# An Efficient Multi-Step Framework for Malware Packing Identification

Jong-Wouk Kim[1], Yang-Sae Moon[1,2], and Mi-Jung Choi[1,2,*]

Department of Computer Science, Kangwon National University[1]

IGP. In Bigdata Convergence, Kangwon National University[2]

{jw.kim, ysmoon, mjchoi}@kangwon.ac.kr

## Abstract

Malware developers use combinations of techniques such as compression, encryption, and obfuscation to bypass anti-virus software. Malware with anti-analysis technologies can bypass AI-based anti-virus software and malware analysis tools. Therefore, classifying pack files is one of the big challenges. Problems arise if the malware classifiers learn packers' features, not those of malware. Training the models with unintended erroneous data turn into poisoning attacks, adversarial attacks, and evasion attacks. Therefore, researchers should consider packing to build appropriate malware classifier models. In this paper, we propose a multi-step framework for classifying and identifying packed samples which is consists of pseudo-optimal feature selection, machine learning-based classifiers, and packer identification steps. In the first step, we use the CART algorithm and the permutation importance to preselect important 20 features. In the second step, each model learns 20 preselected features for classifying the packed files with the highest performance. As a result, the XGBoost, which learned the features preselected by XGBoost with the permutation importance, showed the highest performance than any other experiment scenarios with an accuracy of 99.67%, an F1-Score of 99.46%, and an area under the curve (AUC) of 99.98%. In the third step, we propose a new approach that can identify packers only for samples classified as Well-Known Packed.

1. Introduction

Malware is one of the persistent and serious problems in cybersecurity. Despite advances in security technologies, it is still difficult to detect and defend against malware threats. At the opening ceremony of the 2018 Pyeongchang Winter Olympics, a serious problem occurred that Cisco's Wi-Fi and official website were interrupted. They found out that malware threats cause serious problems through internal investigations [1]. Effective malware detection and analysis have become an essential element for the users and service vendors in order to minimize various problems that can occur due to malware such as personal information leakage, brand reputation degradation, and service quality degradation.

Analysts collect and analyze malware to prevent the spread of malware. However, malware developers also use a variety of anti-analysis techniques such as packing, anti-VM, and anti-debugging to extend the lifespan of malware. Among the many anti-analysis techniques, packing compresses and encrypts the original program and adds anti-analysis techniques that interfere with the analysis. The stub code unpacks the original program when a user executes a packed file. As shown in Figure 1, the packer appends various anti-analysis techniques while compressing the code and data sections of the original program into one or more sections. When the packed file is loaded and executed in memory, stub code unpacks the encrypted data in the same section or another section to execute the original program.

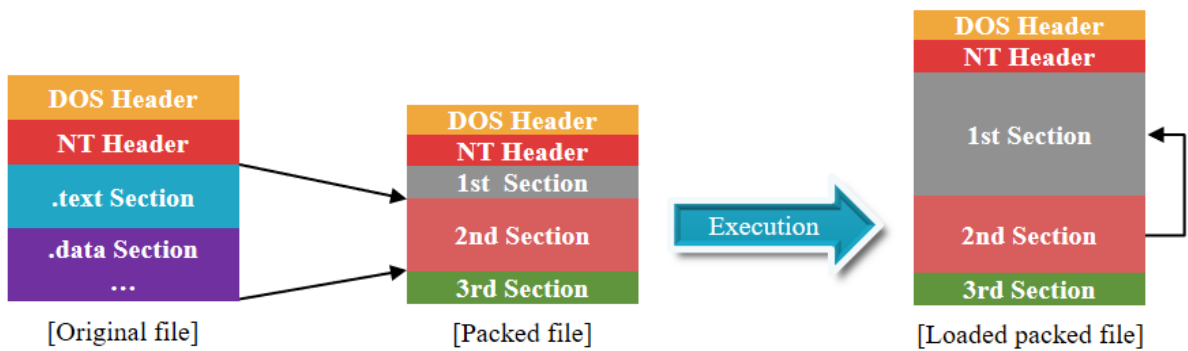

Figure 1. The flow of packing the original program and executing the packed program

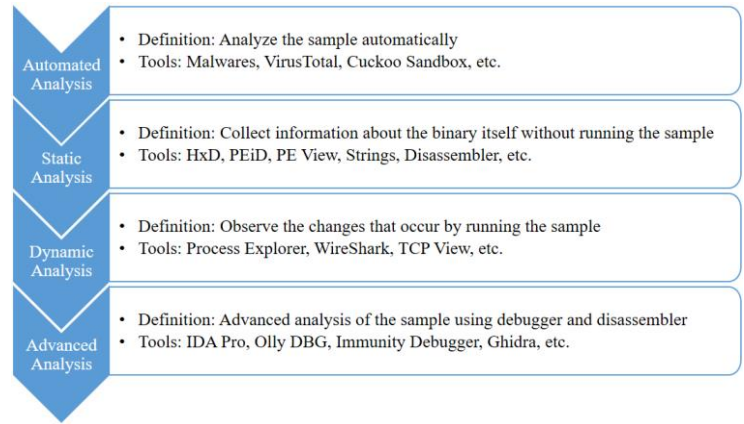

Figure 2. An example workflow for analyzing malware

Analysts analyze malware using automatic, static, dynamic, and advanced analysis, as shown in Figure 2. Automated analysis is one of the analysis methods that uses APIs or automated tools such as sandboxing tools or VirusTotal to discover the flow or type of malware. Static analysis is an analysis method that analyzes the characteristics of strings, headers, and functions by decomposing the malware itself using HxD, PE View, and Strings to obtain the static characteristics and information of the malware. Dynamic analysis is an analysis method that executes malware and observes actual behavior. Analysts use tools such as WireShark and TCP View to monitor network communications, and monitor processes using tools such as Process Explorer. Analysts use debuggers or other tools to observe and analyze the behavior, memory, registers, and registry of malware details in advanced analysis.

However, malware developers bypass antivirus systems and analysis by combining anti-analysis techniques such as packing, anti-VM, and anti-debugging. Researchers can only collect information inserted by the packer instead of the original program if they collect packed malware samples and extract features from them. In dynamic analysis, packed malware can detect and bypass the analysis environment, and the analysis result may differ from the actual malicious behavior. In the advanced analysis, analysts need more time and effort due to anti-analysis techniques such as anti-debugging and anti-emulation of the packed malware. Many researchers consistently study models for detecting malware by combining the sequence of machine language (opcode) and deep learning [2][3]. However, they do not consider encryption such as packing and obfuscation.

Packing not only interferes with the detection and analysis of malware but can also attack machine learning-based malware detection models while training them. Kearns and Li demonstrated a range of maliciously selected errors in the training data of the Valiant's Probably Approach Correct (PAC) training framework, proving that a model must have at least a certain percentage of correct data to function properly [4][5][6][7]. If a poisoning attack occurs while the model is training, the model may be biased incorrectly due to inappropriate and adversarial examples. To prevent such problems and accurately detect malware, it is essential to consider packed samples while the model is training.

In addition, packed files continue to increase every year. Rahbarina et al. surveyed three million samples collected in 2014 and found that 58% of malicious files and 54% of benign files were packed by well-known packers [8]. Morgenstern and Pilz found that 35% of malicious files were packed by custom packers [9]. Pedrero et al. classified packers into six groups and found malware samples that were packed multiple times by custom packers. Among them, the authors found that 65.6% of packers had inconsistent unpacking routines [10].

In this paper, we propose a step-by-step framework that found the pseudo-optimal features, classifying packed malware into the Not Packed, Custom Packing, and Well-known Packed. At last, the framework identifies packers that classified as the Well-known Packed by the model in step 2. This paper presents the following contributions:

(1) We propose a novel multi-step framework consisting of feature selection, packed classification, and packer identification steps.

(2) We present a pseudo-optimal feature selection for packed malware detection exploiting the CART algorithm and the permutation importance with various decision trees.

(3) Using a small number of preselected features, we trained light machine learning models that determine whether a file is packed or not.

(4) Analyzing the large-scale of a well-known packed malware, we classify the type of the packer.

(5) We proved our framework by analyzing and experimenting with various experimental scenarios.

The composition of this paper is as follows. Section 2 introduces malware detection, packing detection, and adversarial attacks on machine learning models. Section 3 describes the multi-step framework proposed in this paper. In the first step, we preselect pseudo-optimal features for detecting and classifying packed malware using the CART algorithm and the permutation importance. The machine learning models classify samples into Not Packed, Custom Packed, and Well-known Packed in the second step. In the third step, the framework can identify the packer of the sample classified as Well-known Packed. Section 4 describes the experiments and evaluates each step in the multi-step framework. XGBoost that learns preselected features by XGBoost and the permutation importance achieved the highest performance among the experimental scenarios and the recent work. In the third step, it identifies the packer of the sample which is classified as the Well-known Packed, and we compared results with an existing tool. Section 5 summarizes the proposed framework and describes the future research.

## 2. Related Work

### 2.1 Malware classification using machine learning

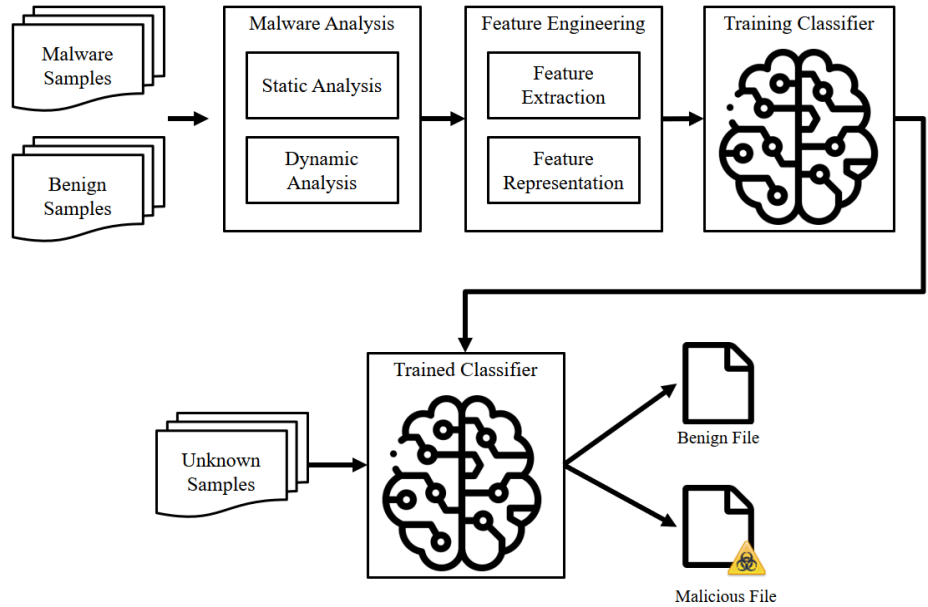

Figure 3. Flowchart of malware detection architecture using machine learning

Machine learning is a useful algorithm for solving classification, clustering, and regression problems. Many researchers try to solve cybersecurity problems using machine learning models. They analyze malware samples, find robust features to build an accurate classifier, and study effective models to end this endless war between attackers and security engineers. Figure 3 is a simplified flow diagram of the malware detection architecture based on a machine learning model. It is essential to find the robust features by analyzing malicious and benign samples, and the model learns these features. We can get these features through the analysis methods shown in Figure 2.

Nataraj et al. proposed a malware classifying method for the first time that visualizes malware binary files as grayscale images and calculates the similarity between the images [11]. The authors achieved an accuracy of 98.08% by performing classification experiments using 25 classes and 9458 malware samples. Many researchers extend the classification method proposed by Nataraj et al. to classify PE-based and Android-based malware samples [12][13][14]. Yan et al. proposed a classification algorithm that integrates PRICoLBP (pairwise rotation invariant co-occurrence local binary pattern) and TF-IDF (term frequency-inverse document frequency) transformation [15]. PRICoLBP-TFIDF achieved higher accuracy against a strong immunity to weak encryption, code segment relocation, redundant data, and instructions. However, strong encryption malware or fragmented malware bypass PRICoLBP-TFIDF.

Many researchers study classifiers that have trained not only gray-scale images but also the order of the opcode sequences. Jeon and Moon proposed the CRNN (convolutional recurrent neural networks) model for detecting malware using opcode sequences [16]. It consists of an opcode-level convolutional autoencoder that compresses long opcode sequences into short sequences at the front, and the DRNN (dynamic recurrent neural networks) that learns and classifies the compressed sequences. However, it showed relatively lower performance compared to other works.

Many researchers study several features through dynamic analysis as well as static analysis. Sihwail et al. proposed a system that applies memory forensics to extract artifacts from memory and combines them with features from dynamic analysis of malware [17]. In addition, the authors achieved high accuracy of 98.5% and low false positives of 1.7% using the pre-modeling technique for feature engineering and the SVM classifier. Xue et al. proposed the Malscore system that can classify malware using probability scoring and machine learning [18]. In the first phase of Malscore, it uses a CNN model for training grayscale images. In the second phase, Malscore learned API call sequences that transformed into *n*-gram. Malscore achieved a high accuracy of 98.8% using probability scores to check the reliability of the classification results.

2.2 Packed file detection using machine learning

Zhang et al. proposed a technique to detect packed malware based on the system calls [19]. The authors extracted the context of the system calls from benign and malicious samples using the sandbox environment for detecting packed malware. They extracted sensitive system call contexts and trained the DBN (deep belief network) model to achieve a high accuracy. Biondi et al. proposed effective, efficient, and robust 119 features to detect and classify packed malware into each packer [20]. The authors built the ground truth by labeling it in three different ways. They combined features and algorithms with more than 1500 scenarios to research which machine learning model and features could show the best performance.

Park et al. proposed a new framework for inferring the lineage of packed malware [21]. The authors used static analysis and dynamic analysis to extract a common feature set and a family feature set, and they appended two steps for identifying the version of packed malware. They used a common feature set to classify packed malware and solved an agglomerative clustering problem using a family feature set extracted from 12221 files. The authors were able to match not-packed and packed files and infer the lineage of packed files using the common feature set and the family feature set. Vasan et al. proposed a new architecture for classifying packed and not-packed malware [22]. This architecture used ensemble CNNs and showed robust performance even in encrypted malware. The architecture also showed low false alarms using 9339 malware images. In the case of not-packed malware, the architecture achieved high accuracy of 99.0%, and in the case of packed malware, it achieved an accuracy of 98.0%. However, it is difficult to say that Park et al. and Basan et al. properly evaluated the models' performance because they used too few samples.

2.3 Adversarial attacks on neural networks

Goodfellow et al. proved that adversarial examples deceive the machine learning models [23]. They also showed that machine learning models inconsistently classify adversarial examples due to the poisoned training data set generated with perturbation. Zügner et al. experimented more demanding addiction and casual attacks using CNNs to focus on the training phase of machine learning models [24]. The authors found that even a little interference significantly reduces classification accuracy. It has also been found that these attacks can deceive many classification models. Thus, they proposed the efficient Nettack that utilizes gradual calculations to cope with attacks that can deceive machine learning models.

Adversarial attacks can be classified into a gradient-based attack, a score-based attack, a decision-based attack, and a transfer-based attack. The gradient-based attack is a kind of white-box attack. The white-box attack assumes that the adversary has access to model parameters on top of being able to get labels for a given input. Goodfellow et al. proposed an FGSM (fast gradient sign method) to explore the gradient direction of a decision boundary [23]. Carlini et al. proposed a powerful attack against defensive distillation and demonstrated their work to evaluate the efficacy of potential defenses. It uses the Adam optimizer to find adversarial examples [42]. It has the disadvantage to generate adversarial examples if the attackers are inaccessible to the model or the details are unknown. Therefore, many researchers are paying attention to various black-box attacks recently. Attackers generate adversarial examples using only queries without any information about a target model. However, many service vendors limit users' queries, and attackers need to reduce the number of queries because sending a large number of queries within a short time is perceived as a scam or a threat.

Score-based attacks make a large number of queries to the target model and exploit the ouput probabilities to generate adversarial examples. Narodytska et al. proposed a local search attack that measures the model sensitivity to individual pixels [36]. The authors proposed black-box attacks by adding perturbation to a randomly selected single pixel then constructing a small set of pixels to perturb by greedy local search. Chen et al. tried to attack the target model by directly estimating the gradients of the target models for generating adversarial examples [37]. They used zeroth order stochastic coordinate descent along with dimension reduction, hierarchical attack and importance sampling techniques to efficiently attack the target model.

Decision-based attacks use only the final output of the target model. Brunner et al. proposed a Biased Boundary Attack that biases the sampling procedure by combining low-frequency random noise with the gradient of an alternative model [38]. The authors combined image frequency, regional masks, and surrogate gradients biases to generate adversarial examples. They evaluated their performance against the ImageNet classifier and the Google Cloud Vision API with just a few hundred queries. Ilyas et al. proposed three black-box threat models that characterize many real-world systems: the query-limited setting, partial-information setting, and the label-only setting [39]. The authors proposed a variant of the NES (natural evolutionary strategies) to generate adversarial examples for query efficiency. The authors also provided theoretical comparisons with previous works about adversarial examples by correlating finite difference methods for the NES and Gaussian bases.

Transfer-based attacks use the prediction of the target model to train the surrogate model for generating adversarial examples. Papernot et al. introduced an attack based on a novel substitute training algorithm using synthetic data generation methods [40]. A novel attack strategy proposed by the authors is to train a substitute model on a synthetic dataset. An attacker uses the output label by the target model as an input to the substitute model. The parameters of the substitute model craft adversarial examples, which are misclassified in the substitute model as well as in the target model.

Liu et al. studied both non-targeted and targeted adversarial examples and showed that while transferable non-targeted adversarial examples are easy to generate, targeted adversarial examples using other researches almost never transfer with their target labels [41]. The authors proposed a novel ensemble-based attack model to generate transferable adversarial examples. They used an image classification service, Clarifai.com, to confirm that adversarial examples were misclassified.

Lit et al. proposed an adversarial machine learning model to detect malware based on opcode $n$-grams [25]. The authors collected 7927 malicious samples and 4070 benign samples, extracted opcode $n$-gram sequences using TF-IDF, and generated adversarial features using XGBoost. SVM, DNN, and XGBoost which learned the original opcode $n$-gram, classified test samples well, but failed to classify the adversarial features properly.

# 3. Multi-Step approach to classify packed malware

## 3.1 Overview of entire framework

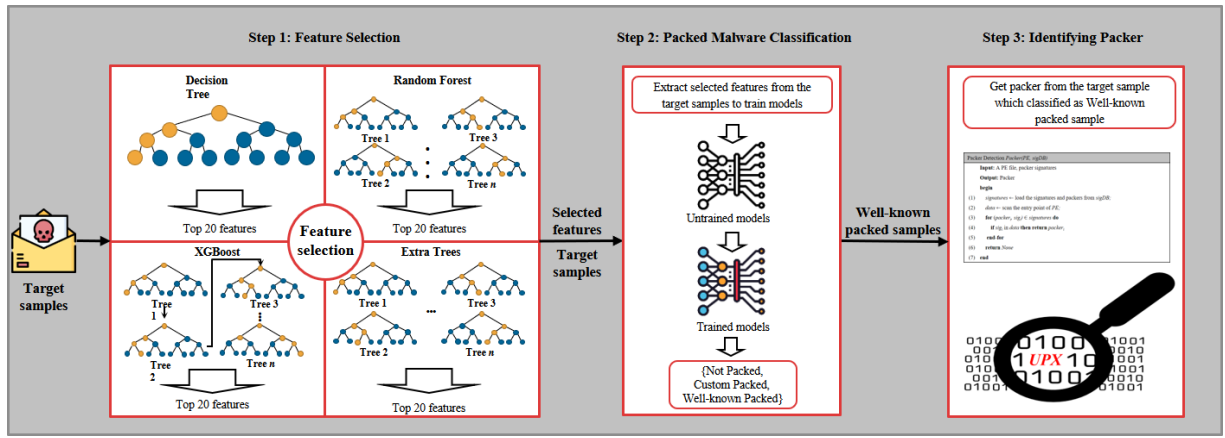

Figure 4. Multi-step approach to classify and identify packed malware

In this paper, we propose a step-by-step framework to classify and identify packed malware. The main goal of the framework is to select pseudo-optimal features that can classify whether malware is packed or not and research effectual model for classification. Finally, in the case of the Well-known Packed samples, the framework recognizes the packer of it.

## 3.2 Step 1: Feature selection

### 3.2.1 CART (classification and regression trees) algorithm

The framework preselect the top 20 features using the CART algorithm that calculates the importance of the features proposed in the recent work [20]. We use Decision Tree, Random Forest, Extra Trees, and XGBoost that use the CART algorithm for measuring important features and preselecting the 20 most important features. The CART algorithm builds a tree to reduce the Gini impurity and selecting appropriate features to classify the samples into homogeneous classes when partitioning the samples [26][27][28].

If the Gini impurity calculated by Equation 1 significantly decreases, it means that it is an important feature when building a tree. The CART algorithm selects a feature that greatly reduces the Gini impurity as the sixth line of Algorithm 1 and generates a node and so on. The stopping conditions of this algorithm are as follows: (1) when there is no remaining data; (2) when all data belonging to a node have the same feature; (3) when a node's data is below a threshold; (4) when the depth of the tree exceeds a predefined value.

Suppose that we have a simple dataset as shown in Table 1. This dataset represents the packing status of samples according to Entropy of EPS, Entropy of .text, Number of standard sections, and Zero size of uninitialized data, and the tree classifies them as Figure 5. When selecting a feature of the root node in Figure 5, must calculate the Gini impurity of each feature as Table 2. Among the four features, Zero size of raw data shows the lowest Gini impurity. Therefore, the root node can split six Not-Packed samples and eight Packed samples based on the Zero size of raw data.

```
Classification and regression tree (CART) algorithm for a single Decision Tree CART (X, F)
        Input: Training dataset X, feature set F
        Output: Tree T
        begin
(1)     A single tree T with a root node;
(2)     if all stopping criteria have been met then
(3)         T has one node with the most common class in X as label;
(4)         return T;
(5)     else
(6)         Find f ∈ F that best split X using the Gini impurity;
(7)         Create label node with f;
(8)         for possible value v of f do
(9)             X= the subset of X that have v = f;
(10)            F = feature set F – the best split feature f;
(11)            CART (X, F);
(12)            Connect the new label node (f, v) to the parent node;
(13)        end-for
(14)    end-if
(15) end
```

Algorithm 1. CART (classification and regression trees) algorithm for a single tree

Table 1. An example dataset to create a single Decision Tree

| Entropy of EPS | Entropy of .text | Number of standard sections | Zero size of raw data | Label |
|---|---|---|---|---|
| Mid | Mid | 0 | True | Not Packed |
| Mid | Mid | 1 | False | Not Packed |
| Low | Mid | 0 | True | Packed |
| High | High | 0 | False | Not Packed |
| High | Low | 0 | True | Packed |
| High | High | 1 | True | Packed |
| Mid | Low | 0 | True | Packed |
| Mid | Mid | 2 | True | Not Packed |
| Low | High | 1 | True | Packed |
| High | High | 1 | True | Packed |
| High | High | 1 | False | Not Packed |
| Low | Low | 0 | False | Packed |
| Low | Mid | 1 | True | Packed |
| High | Low | 0 | False | Not Packed |

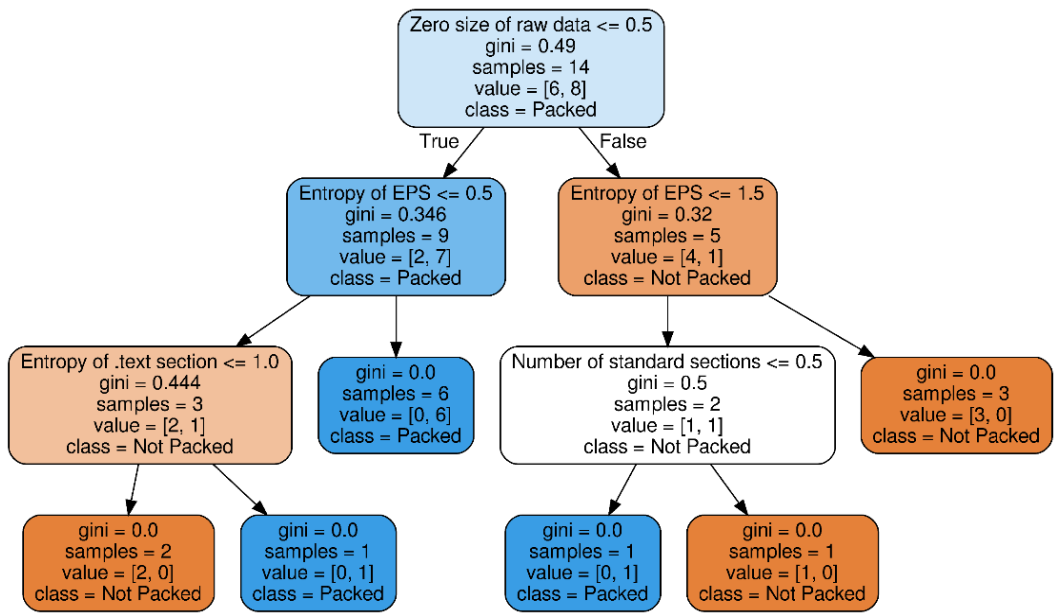

Figure 5. A Decision tree to classify the example dataset

Table 2. The Gini index of each feature to select the one that best splits data at the root node

| Feature | Gini index |
|---|---|
| Entropy of EPS | 0.4642 |
| Entropy of .text | 0.4500 |
| Number of standard sections | 0.4354 |
| Zero size of raw data | 0.3365 |

### 3.2.2 Feature selection based on CART and impurity

```
Algorithm to select important features from one or many trees
      Input: Tree set 𝕋, feature set 𝔽
      Output: The top 20 important features
      begin
(1)       for each tree T_i ∈ 𝕋 do
(2)           for each node j ∈ T_i do
(3)               if j.leftChild = leaf and j.rightChild = leaf then
(4)                   Go to next node;
(5)               else
(6)                   left ← j.leftChild;
(7)                   right ← j.rightChild;
(8)                   nodeImportance_j ← weigthed number of sampels in j × impurity value of j;
(9)                   nodeImportance_j ← nodeImportance_j − weighted number of samples in left × impurity value of left;
(11)                  nodeImportance_j ← nodeImportance_j − weighted number of samples in right × impurity value of right;
(12)                  Go to next node;
(13)          end-for
(14)          for each feature f ∈ 𝔽 do
(15)              featureImportacne_f = (Σ_{k: node k splited by f} nodeImportacne_k / Σ_{j∈all nodes} nodeImportance_j) ÷ Σ_{l∈𝔽} featureImportacne_l;
(16)          end-for
(17)      end-for
(18)      for each feature f ∈ 𝔽 do
(19)          featureImportacne_f = Σ_{m∈𝕋} feautreImportance_f in tree m / Number of trees in 𝕋;
(20)      return The top 20 important features
(21)  end
```

Algorithm 2. Feature selection algorithm from one or many trees

We measure the feature importance of a tree generated by the CART algorithm using Algorithm 2 and Equation 2–4. Algorithm 2 ranks each feature that makes a significant contribution when building a tree. We can measure the importance of each node through Equation 2, calculate the feature importance in the tree using Equation 3, and normalize the feature importance using Equation 4. In the case of calculating the feature importance across the several trees, calculate the average of the feature importance calculated from all trees as in Equation 5.

There are $K$ samples in each class, and $p_i$ represents the probability that sample $i$ belongs to the class. We can get the Gini impurity at the specific node $N_j$ using Equation 1. The node shows low impurity if the node splits the samples heterogeneously. Therefore, it shows low impurity at that node. The tree generated by the CART algorithm reduces the Gini impurity to create nodes and classify data, we can assume that the feature which significantly reduces impurity value is important.

Equation 2–5 can calculate the importance of all features in trees. Equation 2 means the difference between the impurity value of the specific node and the impurity values of its children nodes. It calculates the importance of all nodes in a tree. $I(N_j)$ denotes the importance of the node $N_j$, and $w_j$ means the number of samples in the node $N_j$ and denotes the weight. If the importance value of the specific node is high, the node can be considered as an important node. Equation 3 calculates the importance of each feature in order to divide the sum of all nodes' importance values split by each feature by the sum of all nodes' importance values. We can normalize the importance values through Equation 4 and can use Equation 5 to calculate the feature importance for all trees. The feature importance of Figure 5 is calculated as follows:

- $nodeImportance_{1st}(Zero\ size\ of\ raw\ data) = 14 \times 0.49 - 9 \times 0.346 - 5 \times 0.32 = 2.146$
- $nodeImportance_{2nd}(Entropy\ of\ EPS) = 9 \times 0.346 - 3 \times 0.4444 - 6 \times 0 = 1.794$
- $nodeImportance_{2nd}(Entropy\ of\ EPS) = 5 \times 0.32 - 2 \times 0.5 - 3 \times 0 = 0.6000$
- $nodeImportance_{3rd}(Entropy\ of\ .text\ section) = 3 \times 0.4444 - 2 \times 0 - 1 \times 0 = 1.332$
- $nodeImportance_{3rd}(Number\ of\ standard\ sections) = 2 \times 0.5 - 1 \times 0 - 1 \times 0 = 1.0000$
- $feautreImportance_{Entropy\ of\ EPS} = (2.146 \div 6.872) \div 1 = 0.3484$
- $feautreImportance_{Entropy\ of\ .text\ section} = (1.332 \div 6.872) \div 1 = 0.1934$
- $feautreImportance_{Number\ of\ standard\ sections} = (1.000 \div 6.872) \div 1 = 0.1455$
- $feautreImportance_{Zero\ size\ of\ raw\ data} = (2.146 \div 6.872) \div 1 = 0.3123$

In Figure 5, Entropy of EPS is the most important feature and the second is Zero size of raw data. We build Decision Tree, Extra Trees, Random Forest, and XGBoost that use Algorithm 2 and Equation 1–5 to select the 20 most important features among the 119 features proposed by Biondi et al. [20].

$$G(N_j) = \sum_{i=1}^{K} p_i(1-p_i) = 1 - \sum_{i=1}^{K} p_i^2 \qquad \text{Equation 1}$$

$$I(N_j) = w_j \times G(N_j) - w_{j\_left} \times G(N_{j\_left}) - w_{j\_right} \times G(N_{j\_right}) \qquad \text{Equation 2}$$

$$I(f_{i \in all\ features}) = \frac{\sum_{j:node\ j\ splits\ on\ feature\ i} I(N_j)}{\sum_{k \in all\ nodes} I(N_k)} \qquad \text{Equation 3}$$

$$I(f_{i \in all\ features})^{norm} = \frac{I(f_j)}{\sum_{i \in all\ features} I(f_i)} \qquad \text{Equation 4}$$

$$I(f_{i \in all\ features})^{all\ trees} = \frac{\sum_{t \in all\ trees} I(f_i)^{norm}\ in\ t}{Number\ of\ trees} \qquad \text{Equation 5}$$

### 3.2.3 Permutation Importance

```
Permutation Importance permutationImportance(X, F, M )
    Input: Training dataset X, feature set F, machine learning models M
    Output: Scored feature set
    begin
(1)     P ← 30% of X for permutation importance;
(2)     X ← X – P;
(3)     Train models M with the P;
(4)     for possible feature f of F do
(5)         Shuffle the records of f in the X;
(6)         Measure and record performance;
        end-for
        Measure the scored of feature set;
        return Top 20 importance feature set;
```

Algorithm 3. Permutation importance algorithm.

Permutation importance ranks the importance of all features by calculating how much it affects performance loss when changing or shuffling the data of each feature as Algorithm 3. It scores the influence of each feature by calculating how much performance loss compared to the unshuffled dataset and shuffled dataset. The performance will suffer if the classifier heavily relies on the specific feature to be shuffled. Because the shuffling procedure breaks the relationship between the feature and the target.

It is a kind of model inspection technique that can be used for any model when the data is tabular.

This technique has the advantage that it is model agnostic and can measure multiple times using different permutations. It also has disadvantages. The results of permutation importance are always different. However, increasing the number of shuffle steps can reduce the variance of errors by augmenting the computations. Also, random shuffling of features is likely to produce very unrealistic data combinations. This is easy to happen when the correlations between features are high. Increasing the uncertainty of the data could have a significant impact on the predictions. Showing high importance may not be the actual feature importance we want.

3.3 Step 2: Light machine learning models for classifying packed malware

```
Implying Machine Learning Classifier Cls(X, F, Y, M)
    Input: Training dataset X, pre-selected features F, label Y, machine learning models M
    Output: Trained machine learning models T
    begin
(1)     for each model M ∈ M do
(2)         Classifier ← supervised learning by model M with X, F, and Y;
(3)         Append Classifier in T;
(4)     end-for
(5)     return Best Classifier in T
(6) end
```

Algorithm 4. Training machine learning models

In step 2, the 10 models learn the important 20 features selected by step 1. We use several classification algorithms implemented in the *scikit-learn* [29] and the *Keras* [30] to find the model that shows the best performance for classifying packed malware into three classes. If we classify packed samples into each packer, the models misclassify well-known packed samples that have a small number of samples. Therefore, we classify malware into Not Packed, Custom Packed, and Well-known Packed. Every model learns the training dataset $\mathbb{X}$, the preselected features $\mathbb{F}$, and the labels $\mathbb{Y}$, returns the models to find the best performing features and the as Algorithm 4.

All models $\mathbb{M}$ perform supervised learning. Supervised learning is one of the learning methods that infer a function or decision boundary from training data. Each model learns preselected features from step 1 and ground truth. After training, evaluate the models using test data, and step 2 returns the model that show the highest performance.

## 3.4 Step 3: Identify the packer using signatures

Step 3 of the proposed framework identifies the packer of a packed sample that is classified as Well-known Packed in step 2. We use the *pefile* [31] for detecting the packer signature from the PE-based malware as Algorithm 5. It confirms the EPS of the binary file for identifying the packer signature. If several signatures stored in the database exist in the EPS, it returns the last detected packer signature.

The signature database has 4200 signatures provided by the *PEiD* [32]. Figure 6 shows some of the packer signatures stored in the database. The first line in Figure 6 represents the packer and its version information, and the second line indicates the packer's signature in hexadecimal. The 'ep_only' attribute indicates whether a signature can be found in the entry point section as True or False.

```
[ASPack 1.05b by]
signature = 75 00 E9
ep_only = true

[ASPAck 1.061b]
signature = 90 90 75 00 E9
ep_only = true

[ASPack 1.08]
signature = 90 90 90 75 01 90 E9
ep_only = true

[UPX -> www.upx.sourceforge.net]
signature = 60 BE ?? ?? ?? 00 8D BE ?? ?? ?? FF
ep_only = true

[UPX 0.50 - 0.70]
signature = 60 E8 00 00 00 00 58 83 E8 3D
ep_only = true

[yoda's Crypter 1.3 -> Ashkbiz Danehkar]
signature = 55 8B EC 53 56 57 60 E8 00 00 00 00 5D 81 ED 6C 28 40 00 B9 5D 34 40 00
ep_only = true
```

Figure 6. Examples of packer signatures stored in the signature database.

```
Packer Detection Packer(PE, sigDB)
    Input: A PE file, packer signatures
    Output: Packer
    begin
(1)   signatures ← load the signatures and packers from sigDB;
(2)   data ← scan the entry point of PE;
(3)   for (packer_i, sig_i) ∈ signatures do
(4)     if sig_i in data then return packer_i
(5)   end-for
(6)   return None
(7) end
```

Algorithm 5. Algorithm to identify the packer

## 4. Experimental results and discussion

### 4.1 Experimental setup

```
Labeling algorithm  Labeling(A PE file)
    Input: A PE file
    Output: Not Packed or Custom Packed or Well-known Packed
    begin
(1)     if No EP Section then return Custom Packed;
(2)     else if Packer Signature Found then return Well-known Packed;
(4)     else if "WRITE" characteristic and EPS Entropy ∈ Packing-Range then
(5)         return Custom Packed;
(6)     else return Not Packed;
(7)     end if
(8) end
```

Algorithm 6. Labeling Algorithm

We collected a total of 214001 malware samples from VirusShare[1] and labeled them using a part of the framework proposed in the previous study as Algorithm 6 [33]. We labeled 87502 samples as Custom Packed that have hidden EPS because packers can hide a proper address of the entry point while packing binary files. In the case that samples have an appropriate entry point and has the write property, and the entropy value of the EPS belongs to the Packing-Range suggested in the previous study, we labeled them as also Custom Packed. Packed binary files need the write attribute to restore the IAT (import address table), original code, and data. We labeled 25955 samples as Well-known Packed that have findable EPS and packer signatures. Otherwise, we labeled 100544 samples as Not-Packed.

Figure 7 shows the packer distribution of packed malware. UPX is the most frequent with 6748 samples, followed by Netopsystems with 4482 samples, and 35 packed malware samples each have a unique packer. Appendix 1 shows detailed information about Well-known Packed samples. We used 214001 PE-based malware samples and split the dataset in a 70–30 ratio for training and testing. Table 3 shows the composition of the dataset for the experiments.

---

[1] www.virusshare.com

Table 3. Distribution of train and test set

| Dataset | Group | | | Total |
|---|---|---|---|---|
| | Custom Packed | Well-known Packed | Not Packed | |
| Train Set | 61251 (28.6%) | 18168 (8.5%) | 70380 (32.9%) | 149799 (70.0%) |
| Test Set | 26251 (12.3%) | 7787 (3.6%) | 30164 (14.1%) | 64202 (30.0%) |

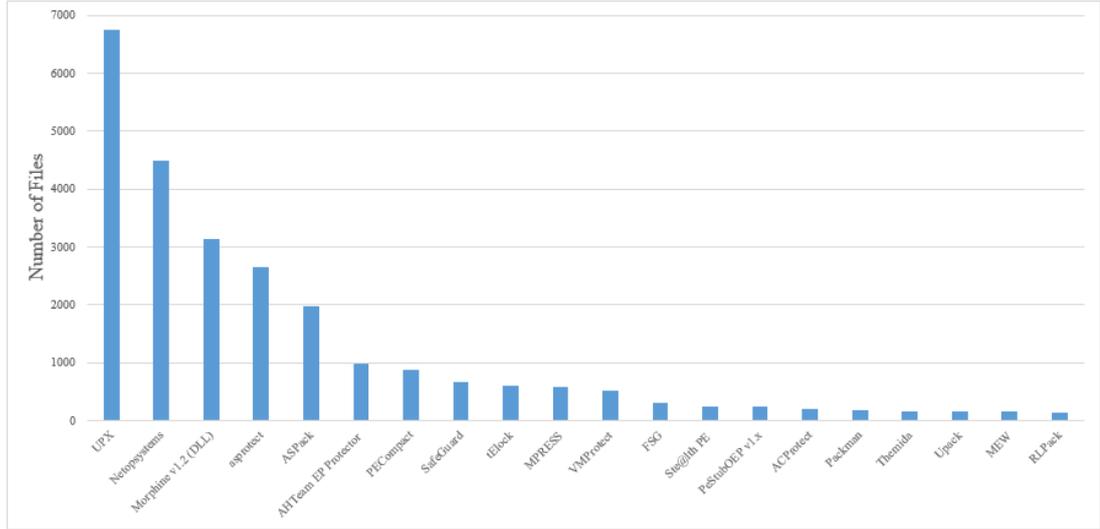

Figure 7. Distribution of well-known packers (top 20)

We compared an accuracy, F1-Score, AUC (area under the ROC curve) widely used by many researchers for detailed evaluation [34][35]. We evaluated the machine learning models the following factors in step 2:

- True Positives (TP): The points classified as positive by the model that are actually positive.
- True Negatives (TN): The points classified as negative by the model that are actually negative.
- False Positives (FP): The points classified as positive by the model that are actually negative.
- False Negatives (FN): The points classified as negative by the model that are actually positive.
- Precision: The number of true positives divided by the number of true positives and the number of false positives ($\frac{TP}{TP+FP}$).

- True Positive Rate (TPR), Recall: The number of true positives divided by the number of true positives and the number of false negatives ($\frac{TP}{TP+FN}$).
- False Positive Rate (FPR): The number of false positive divided by the number of false positive and true negatives ($\frac{FP}{FP+TN}$).

We calculated an accuracy using Equation 6 as the proportion of samples with correct answers among all predictions. we also measured the F1-score and AUC to avoid an accuracy paradox. F1-score is the harmonic mean of precision and recall as Equation 7, and it takes both false positives and false negatives to measure the performance in the case of using imbalanced data. AUC is the area under the ROC curves, and it represents the model's ability to classify each class. ROC curves show the performance of a classification model at all classification thresholds with TPR and FPR.

$$\text{Accuracy} = \frac{TP + TN}{TP + TN + FP + FN} \qquad \text{Equation 6}$$

$$\text{F1-score} = 2 \times \frac{Precision \ \times Recall}{Precision + Recall} \qquad \text{Equation 7}$$

## 4.2 Experimental results

### 4.2.1 Step 1: Comparing feature importance scores

We preselected 20 pseudo-optimal features among 119 features using 214001 malware samples and Decision Tree, Extra Trees, Random Forest, and XGBoost. In the case of Random Forest, Extra Tree, and XGBoost, we built 100 trees to preselect important features. The features selected by each model show slightly different results depending on the model as shown in Figure 8. However, the Entropy of the entry point section has the highest importance values in most models.

As shown in Figure 8(a) and Figure 8(e), Decision Tree measured the importance of the Entropy of entry point section very high compared to other models. This is because the models that build multiple trees calculate the average of the feature importance in the 19th line of Algorithm 2. In other words, the models that build a large number of trees measure the feature importance more carefully because they consider various features in multiple trees.

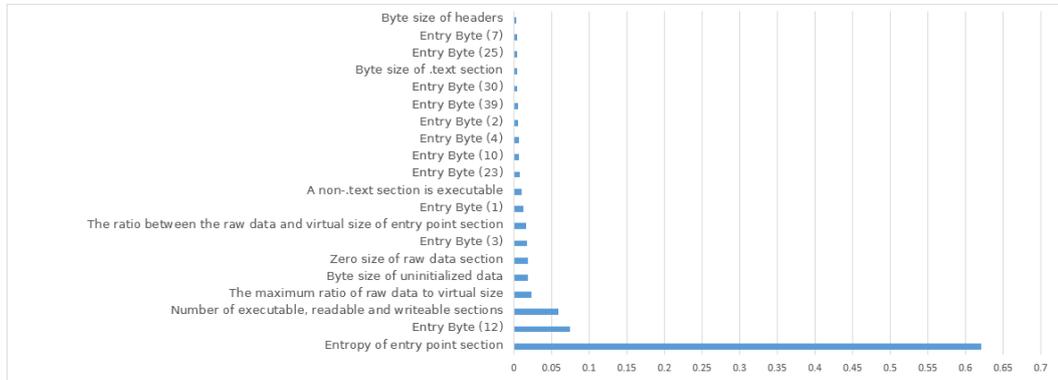

(a) Feature importance scores of CART with Decision Tree

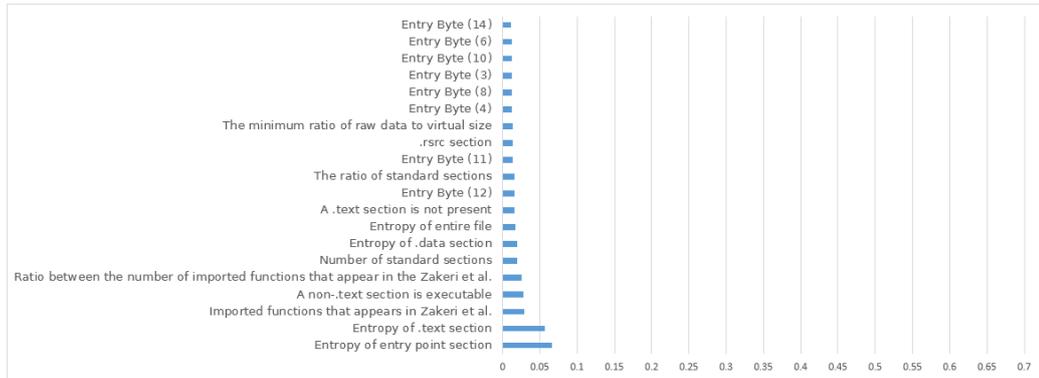

(b) Feature importance scores of CART with Extra Trees

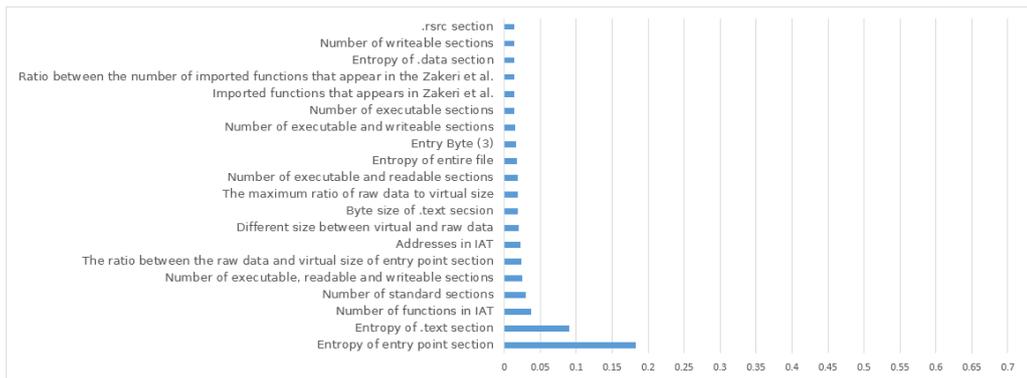

(c) Feature importance scores of CART with Random Forest

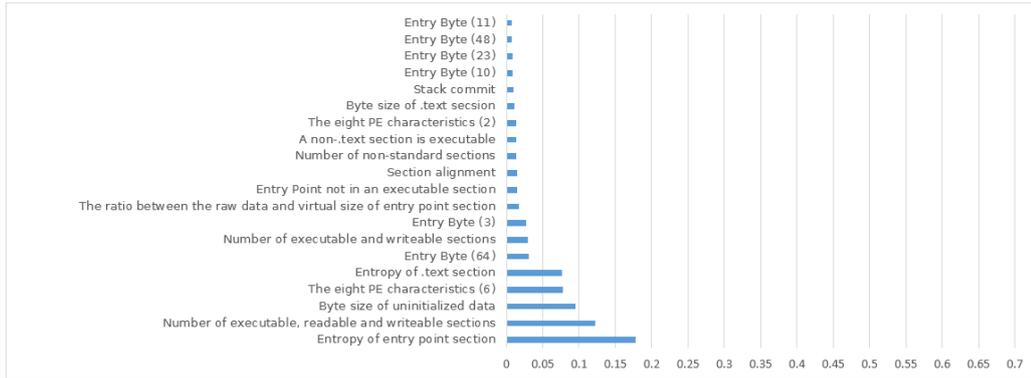

(d) Feature importance scores of CART with XGBoost

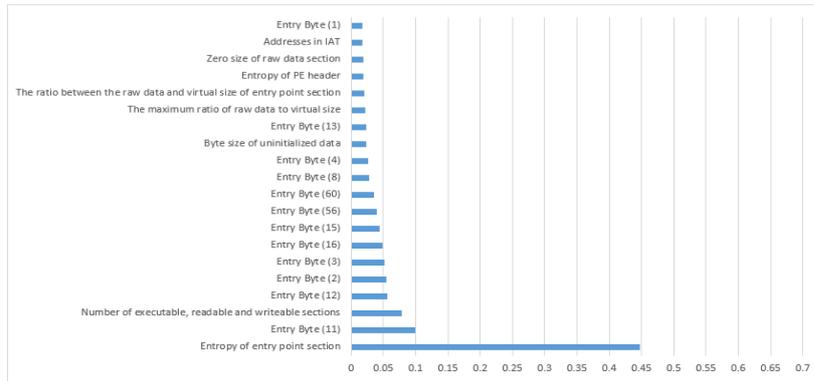

(e) Permutation importance scores with Decision Tree

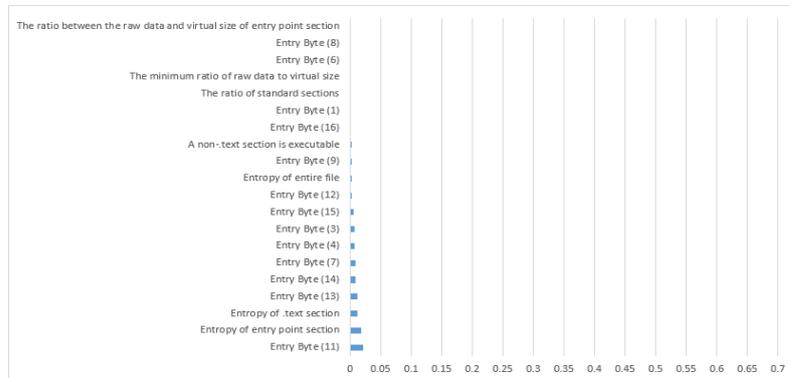

(f) Permutation importance scores with Extra Trees

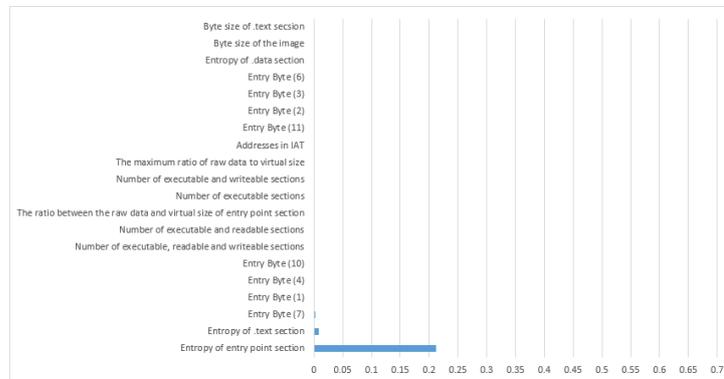

(g) Permutation importance scores with Random Forest

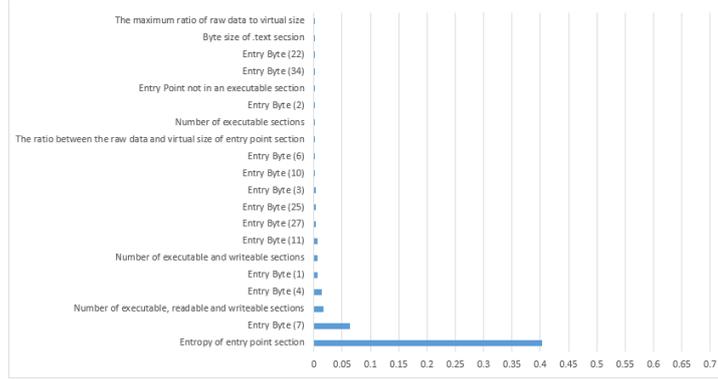

(h) Permutation importance scores with XGBoost

Figure 8. Top 20 important features selected by various machine learning models

### 4.2.2 Step 2: Comparing machine learning models

Table 4 shows the results of comparing various experimental scenarios including the recent work. It shows which scenario is the best by comparing the accuracy, F1-Score, and AUC. Among the many experimental scenarios, the XGBoost that learned the top 20 most important features selected by XGBoost with the permutation importance performed the best. It has shown excellent performance because it provides a parallel tree boosting that solves many problems quickly and accurately.

Figure 9 shows the confusion matrix of some cases. Figure 9(a) shows the best performance of all experiment scenarios. It classified the Not-Packed, and Custom-Packed samples perfectly and the Well-known Packed samples more perfectly than prior work [20]. This is because the Well-Known Packed samples were the fewest. Nevertheless, Figure 9(a) shows the highest performance among the experimental scenarios including the recent study.

Table 4. Classifying results of each feature selection model and each classifying model

| Feature Selection Model | Classifying Model | Accuracy | F1-Score | AUC |
|---|---|---|---|---|
| Decision Tree (CART) | Linear SVM | 0.2981 | 0.2772 | 0.5195 |
| | SVM | 0.7113 | 0.6457 | 0.7876 |
| | Logistic Regression | 0.4654 | 0.2254 | 0.6915 |
| | Naïve Bayes | 0.6884 | 0.6364 | 0.8536 |
| | kNN (*k*=3) | 0.9047 | 0.9032 | 0.9633 |
| | MLP | 0.7011 | 0.6991 | 0.8072 |
| | Decision Tree | 0.9893 | 0.9829 | 0.9898 |
| | Random Forest | 0.9943 | 0.9911 | 0.9998 |
| | Extra Trees | 0.9868 | 0.9844 | 0.9994 |
| | XGBoost | 0.9954 | 0.9927 | 0.9998 |
| Extra Tree (CART) | Linear SVM | 0.4610 | 0.3825 | 0.6600 |
| | SVM | 0.5844 | 0.5595 | 0.5371 |
| | Logistic Regression | 0.7793 | 0.7367 | 0.9019 |
| | Naïve Bayes | 0.6529 | 0.6092 | 0.8397 |
| | kNN (*k*=3) | 0.9433 | 0.9444 | 0.9787 |
| | MLP | 0.8999 | 0.8999 | 0.9531 |
| | Decision Tree | 0.9837 | 0.9747 | 0.9823 |
| | Random Forest | 0.9888 | 0.9833 | 0.9990 |
| | Extra Trees | 0.9853 | 0.9799 | 0.9981 |
| | XGBoost | 0.9887 | 0.9827 | 0.9993 |
| Random Forest (CART) | Linear SVM | 0.4294 | 0.4006 | 0.5669 |
| | SVM | 0.6389 | 0.4696 | 0.7478 |
| | Logistic Regression | 0.4850 | 0.2727 | 0.7024 |
| | Naïve Bayes | 0.7101 | 0.6635 | 0.8608 |
| | kNN (*k*=3) | 0.8708 | 0.8603 | 0.9474 |
| | MLP | 0.6557 | 0.6135 | 0.8322 |
| | Decision Tree | 0.9792 | 0.9643 | 0.9845 |
| | Random Forest | 0.9879 | 0.9791 | 0.9991 |
| | Extra Trees | 0.9848 | 0.9768 | 0.9987 |
| | XGBoost | 0.9868 | 0.9767 | 0.9991 |
| XGBoost (CART) | Linear SVM | 0.6055 | 0.4803 | 0.6480 |
| | SVM | 0.5245 | 0.4800 | 0.5975 |
| | Logistic Regression | 0.4793 | 0.2167 | 0.4987 |
| | Naïve Bayes | 0.6783 | 0.6471 | 0.8436 |
| | kNN (*k*=3) | 0.9060 | 0.9020 | 0.9551 |
| | MLP | 0.6660 | 0.6659 | 0.8281 |
| | Decision Tree | 0.9851 | 0.9745 | 0.9889 |
| | Random Forest | 0.9904 | 0.9836 | 0.9994 |
| | Extra Trees | 0.9876 | 0.9811 | 0.9988 |
| | XGBoost | 0.9902 | 0.9830 | 0.9995 |
| | Linear SVM | 0.4368 | 0.3681 | 0.5845 |
| | SVM | 0.6407 | 0.5761 | 0.7251 |
| Decision Tree (Permutation importance) | Logistic Regression | 0.5824 | 0.3896 | 0.7216 |
| | Naïve Bayes | 0.7167 | 0.6777 | 0.8458 |
| | kNN ($k=3$) | 0.9348 | 0.9329 | 0.9753 |
| | MLP | 0.8251 | 0.8245 | 0.9379 |
| | Decision Tree | 0.9865 | 0.9780 | 0.9899 |
| | Random Forest | 0.9916 | 0.9864 | 0.9998 |
| | Extra Trees | 0.9840 | 0.9808 | 0.9993 |
| | XGBoost | 0.9911 | 0.9861 | 0.9996 |
| Extra Tree (Permutation importance) | Linear SVM | 0.2874 | 0.2657 | 0.5250 |
| | SVM | 035605 | 0.4803 | 0.6386 |
| | Logistic Regression | 0.7000 | 0.6784 | 0.8581 |
| | Naïve Bayes | 0.6803 | 0.6410 | 0.8408 |
| | kNN ($k=3$) | 0.9444 | 0.9461 | 0.9798 |
| | MLP | 0.8880 | 0.8884 | 0.9646 |
| | Decision Tree | 0.9863 | 0.9800 | 0.9897 |
| | Random Forest | 0.9915 | 0.9883 | 0.9995 |
| | Extra Trees | 0.9864 | 0.9832 | 0.9993 |
| | XGBoost | 0.9920 | 0.9892 | 0.9995 |
| Random Forest (Permutation importance) | Linear SVM | 0.5031 | 0.4789 | 0.4821 |
| | SVM | 0.5445 | 0.5219 | 0.6077 |
| | Logistic Regression | 0.5873 | 0.3886 | 0.5571 |
| | Naïve Bayes | 0.6585 | 0.6183 | 0.8109 |
| | kNN ($k=3$) | 0.8760 | 0.8663 | 0.9504 |
| | MLP | 0.7104 | 0.7083 | 0.7666 |
| | Decision Tree | 0.9891 | 0.9818 | 0.9992 |
| | Random Forest | 0.9933 | 0.9888 | 0.9996 |
| | Extra Trees | 0.9899 | 0.9857 | 0.9995 |
| | XGBoost | 0.9943 | 0.9904 | 0.9996 |
| XGBoost (Permutation importance) | Linear SVM | 0.4613 | 0.4006 | 0.5669 |
| | SVM | 0.5571 | 0.4908 | 0.4319 |
| | Logistic Regression | 0.6015 | 0.5904 | 0.6480 |
| | Naïve Bayes | 0.6582 | 0.6398 | 0.7983 |
| | kNN ($k=3$) | 0.9042 | 0.8988 | 0.9498 |
| | MLP | 0.7946 | 0.7899 | 0.8819 |
| | Decision Tree | 0.9913 | 0.9858 | 0.9993 |
| | Random Forest | 0.9945 | 0.9911 | 0.9996 |

|  | Extra Trees | 0.9878 | 0.9859 | 0.9995 |
|---|---|---|---|---|
|  | **XGBoost** | **0.9967** | **0.9946** | **0.9998** |
| Biondi et al.[20] | Decision Tree | 0.9891 | 0.9812 | 0.8787 |
|  | Random Forest | 0.9928 | 0.9885 | 0.9998 |

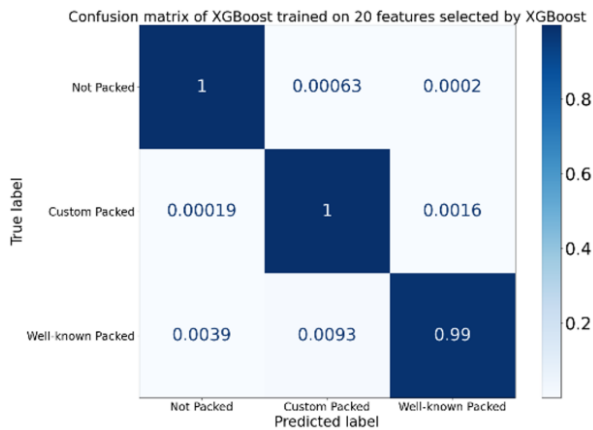

(a) XGBoost classifier that learned the preselected features by the permutation importance and XGBoost

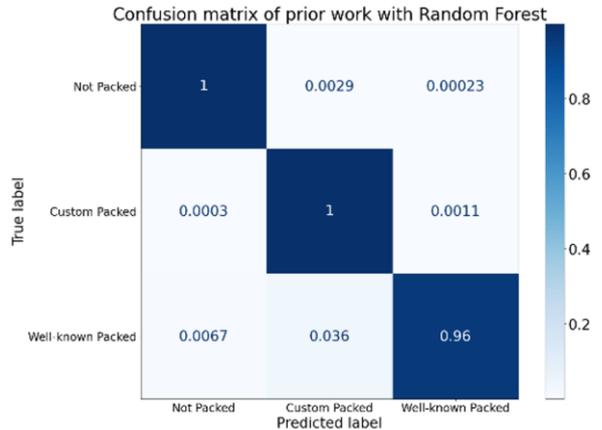

(b) Prior work by Biondi et al. [20]

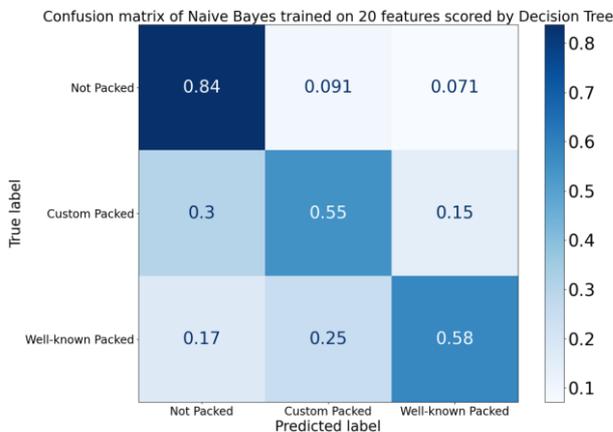

(c) Naïve Bayes classifier that learned the preselected features by the CART and Decision Tree

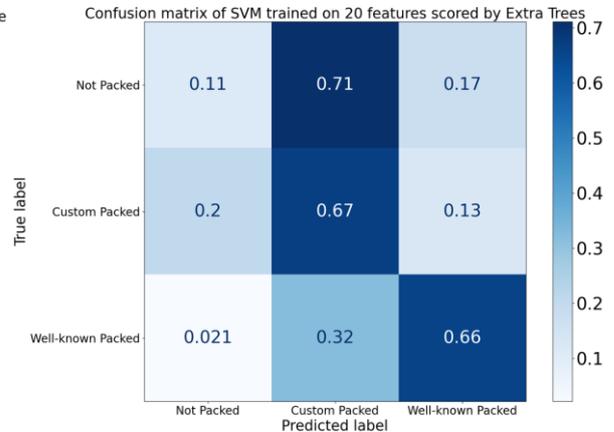

(d) SVM classifier that learned the preselected features by CART and Extra Trees.

Figure 9. The confusion matrix of some cases: (a) XGBoost classifier that learned the preselected features by the permutation importance and XGBoost; (b) Prior work by Biondi et al. [20]; (c) Naïve Bayes classifier that learned the preselected features by the CART and Decision Tree; (d) SVM classifier that learned the preselected features by CART and Extra Trees.

### 4.2.3 Step 3: Identifying packers

The proposed framework identifies the packer in step 3 using *pefile* [31], a multi-platform Python module that analyzes PE files. We identified the packers for 7685 samples classified by the best performing model and compared the results identified by *PEiD* [32] with the results from step 3. We confirmed that the results were in the perfect match. This is because the framework uses the same packer database as PEiD. Figure 10 shows the top 20 frequent packers identified in step 3.

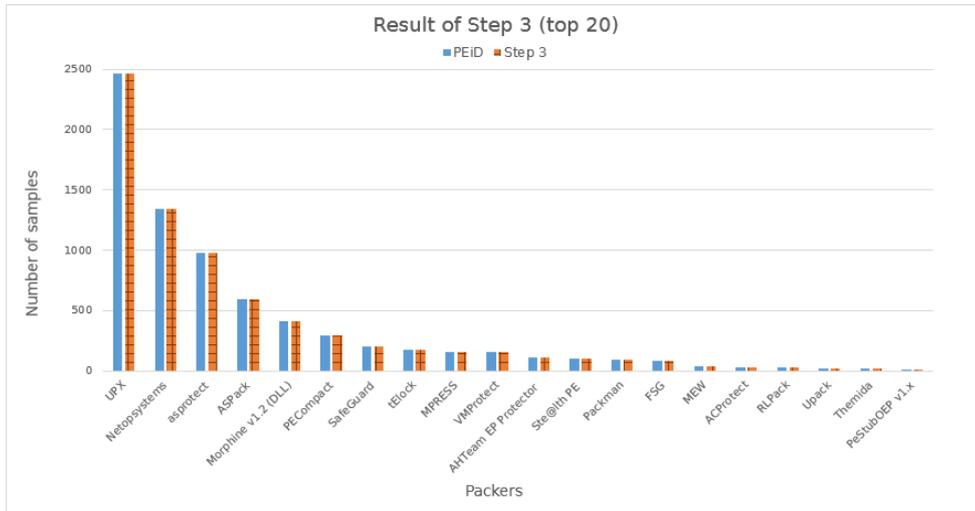

Figure 10. Results of comparing *PEiD* and Step 3

## 5. Conclusion

In this paper, we propose the multi-step framework that can detect whether malware is packed or not using static analysis and machine learning. We preselect 20 important features from 119 features proposed in a recent study through the feature selection algorithms. We also classify packed malware into three classes by comparing the accuracy, F1-Score, and AUC using different feature groups and 10 different machine learning models.

In step 1, we used Decision Tree, Random Forest, Extra Trees, and XGBoost to rank the features that each model considers important. We preselected important 20 features that split the data when building a tree based on the Gini impurity and the permutation importance. As a result, the XGBoost that learned the important 20 features preselected by XGBoost with the permutation importance showed the best performance than any other model including a recent study. When selecting important features, creating a large number of trees showed better results than using only one tree in step 1. The proposed framework showed an accuracy of 99.27%, an F1-Score of 98.84%, and an AUC of 99.96% to classify into three classes in step 2. In the last step, the framework identified the packer of the malware samples classified as Well-known Packed. Therefore, this paper proposed the framework to classify and identify packed malware that is difficult to detect with machine learning—in the case of packers with insufficient data—using the previously used packer detector.

Our approach classifies and identifies the packed files whether the target files are malicious or not. Researchers must consider packing and obfuscation techniques in order to study deep learning models to detect and classify malware. The step-by-step framework proposed in this paper classifies packed files into Well-known Packed, Custom Packed, and Not Packed. The framework identifies the packers that are classified as Well-known Packed samples. We plan to expand this framework that can classify whether the target file is encrypted as well as whether it is malicious or not.

Appendix 1. Distribution of packers

| Packer | # of files | Packer | # of files |
|---|---|---|---|
| UPX | 6,748 | PECompact | 878 |
| Netopsystems | 4,482 | SafeGuard | 675 |
| Morphine v1.2 (DLL) | 3,131 | tElock | 596 |
| asprotect | 2,645 | MPRESS | 571 |
| ASPack | 1,981 | VMProtect | 516 |
| AHTeam EP Protector | 987 | FSG | 297 |
| Ste@lth PE | 250 | kryptor 6 | 132 |
| PeStubOEP | 240 | Enigma | 93 |
| ACProtect | 200 | NsPack | 85 |
| Packman | 169 | Armadillo | 74 |
| Themida | 164 | Petite | 57 |
| Upack | 161 | Fish Pe Packer | 42 |
| MEW | 158 | InstallShield | 38 |
| RLPack | 145 | 캬 v1.0 -> bbb | 33 |
| PE Diminisher | 29 | Hide&Protect | 13 |
| ExeShield | 26 | W32.Jeefo | 12 |
| Safengine Shielden | 19 | Crunch/PE | 12 |
| MoleBox | 18 | HuiGeZi | 11 |
| ZProtect | 15 | NeoLite v2.0 | 11 |
| PKLITE32 | 13 | UNKNOWN | 10 |
| PESpin | 10 | Obsidium | 7 |
| PE Pack | 10 | Embed PE | 7 |
| eXPressor | 9 | PEEncrypt | 6 |
| CreateInstall | 9 | Crinkler | 6 |
| KGB SFX | 8 | Cexe | 6 |
| PEBundle | 7 | Program Protector | 5 |
| PC Guard for Win32 | 5 | PEQuake | 5 |
| EXE Cryptor | 5 | yoda's Protector | 3 |
| ZealPack | 4 | yoda's Crypter | 6 |
| WWPack | 4 | Krypton | 4 |
| kkrunchy | 4 | PECrc32 0.88 | 3 |
| GP-Install | 4 | Feokt | 3 |
| XPack V0.98 | 2 | EZIP | 3 |
| VIRUS - I-Worm.KLEZ | 2 | CRYPToCRACk's PE Protector | 3 |
| Silicon Realms Install Stub | 2 | SVKP | 3 |
| PE Crypt32 | 2 | dUP2 | 2 |
| Pack Master | 2 | Crypto-Lock | 2 |
| North Star PE Shrinker | 2 | Blade Joiner | 2 |
| GameGuard | 2 | ANDpakk2 | 2 |
| EXEStealth | 2 | XComp | 2 |
| EXE Shield | 2 | SafeDisc/SafeCast | 2 |
| BeRoEXEPacker | 2 | nPack | 2 |
| XJ / XPAL -> LiNSoN | 1 | Stone's PE Encrypter | 1 |
| Virogen Crypt v0.75 | 1 | Spalsher | 1 |
| AHPack | 1 | Software Compress | 1 |
| VBOX v4.2 MTE | 1 | SoftSentry | 1 |
| Unnamed Scrambler | 1 | SimplePack | 1 |
| SuperDAT | 1 | Sality.Q | 1 |
| RPoly crypt | 1 | Reg2Exe 2.24 | 1 |
| RCryptor | 1 | PUNiSHER | 1 |
| PeX v0.99 | 1 | PE-Armor | 1 |
| NTPacker | 1 | JDPack | 1 |
| KByS | 1 | iPBProtect v0.1.3 | 1 |
| HPA | 1 | hying's PEArmor | 1 |
| HA Archive | 1 | Freshbind | 1 |
| Gleam | 1 | ExeJoiner | 1 |
| eXcalibur | 1 | EXE32Pack | 1 |
| DBPE vx.xx | 1 | D1S1G | 1 |
| BlackEnergy DDoS Bot Crypter | 1 | Crunch/PE | 1 |
| * PseudoSigner | 1 | Total Well-known Packed malware | 25,955 |